\documentstyle[12pt]{article}
\def\vev#1{\langle #1 \rangle }
\newcommand{\toright}[1]{\smash{
   \mathop{\hbox to 1.3cm {\rightarrowfill}}
    \limits^{#1}}}
\newcommand{\tbtmtrx}[4]{\left(\begin{array}{cc}
            #1 & #2 \\
            #3 & #4
        \end{array}\right)}
\def\Membar {\overline{M2}}
\newcommand{\Dbar}[1]{\overline{D#1}}
\def\Tr {\mbox{Tr}}
\def\half {\frac{1}{2}}
\def\quarter {\frac{1}{4}}
\def\np    { Nucl. Phys. }
\def\pr    { Phys. Rev. }
\def\pl    { Phys. Lett. }
\def\cmp   { Commun. Math. Phys. }

\def\prl   { Phys. Rev. Lett. }

\def\del {\partial}

\def\be{\begin{equation}}     
\def\ee{\end{equation}}
\def\bea{\begin{eqnarray}}     
\def\eea{\end{eqnarray}}
\def\&{&\!\!\!\!\!\!\!\! &}

\def\nn{\nonumber}

\renewenvironment{thebibliography}{\pagebreak[3]\par\vspace{0.6em}
\begin{flushleft}{\large \bf References}\end{flushleft}
\vspace{-1.0em}

\begin{enumerate}\if@twocolumn\baselineskip=0.6em\itemsep -0.2em
\else\itemsep -0.2em\fi\labelsep 0.1em}{\end{enumerate}}
\begin{document}
\baselineskip=0.65cm

%%%%%%%%%%%%%%%%%%%%%%%%%%%%%%%%%%%%%%

%Tachyon Condensation and Graviton Production 
%in Matrix Theory

%%%%%%%%%%%%%%%%%%%%%%%%%%%%%%%%%%%%%%%%%%%%%%%%%%%%%%%%%%

				%titlepage

%%%%%%%%%%%%%%%%%%%%%%%%%%%%%%%%%%%%%%%%%%%%%%%%%%%%%%%%%%
\begin{titlepage}

    \begin{normalsize}
     \begin{flushright}
                 YITP-99-11\\
                 hep-th/9902158\\
                 February 1999
     \end{flushright}
    \end{normalsize}
    \begin{LARGE}
       \vspace{1cm}
       \begin{center}
        Tachyon Condensation and Graviton Production \\
          in \\
        Matrix Theory \\
       \end{center}
    \end{LARGE}
  \vspace{5mm}

\begin{center}
           Hidetoshi Awata,
           \footnote{E-mail :
              awata@yukawa.kyoto-u.ac.jp}
            Shinji Hirano
           \footnote{E-mail :
              hirano@yukawa.kyoto-u.ac.jp}
           and Yoshifumi Hyakutake 
           \footnote{E-mail :
              hyaku@yukawa.kyoto-u.ac.jp}  \\
      \vspace{4mm}
        {\it Yukawa Institute for Theoretical Physics} \\
        {\it Kyoto University}\\
        {\it Sakyo-ku, Kyoto 606-8502, Japan}\\
      \vspace{1cm}

    \begin{large} ABSTRACT \end{large}
        \par

\end{center} 
\begin{quote}
 \begin{normalsize}
We study a membrane -- anti-membrane system in Matrix theory. It in
fact exhibits the tachyon instability. By suitably representing this
configuration, we obtain a (2+1)-dimensional $U(2)$ gauge theory with
a 't Hooft's twisted boundary condition. We identify the tachyon field
with a certain off-diagonal element of the gauge fields in this
model. Taking into account the boundary conditions carefully, we can
find vortex solutions which saturate the Bogomol'nyi-type bound and
manifest the tachyon condensation. We show that they can be
interpreted as gravitons in Matrix theory. 
 \end{normalsize}
\end{quote}

\end{titlepage}
\vfil\eject
%%%%%%%%%%%%%%%%%%%%%%%%%%%%%%%%%%%%%%%%%%%%%%%%%%%%%%%%%%%%%%%%%%%%%%%%%%%%%%

%\setcounter{footnote}{0}

%%%%%%%%%%%%%%%%%%%%%%%%%%%%%%%%%%%%%%%%%%%%%%%%%%%%%%%%%%%%%%%%%%%%%%%%%%%%%%

%Text

%%%%%%%%%%%%%%%%%%%%%%%%%%%%%%%%%%%%%%%%%%%%%%%%%%%%%%%%%%%%%%%%%%%%%%%%%%%%%%

\section{Introduction}

It is well-known that a tachyon develops when a D-brane and an anti-D-brane come up to some critical interval of order of string scale \cite{Green,BS,LP}. Hence the system of a brane and an anti-brane is unstable, and the tachyon fields roll down to a stable point \cite{Sen2}. Recently Sen argued that the D-brane -- anti-D-brane pair is annihilated via the tachyon condensation and in this process certain stable D-branes are produced as soliton solutions on this system \cite{Sen1,Sen3,Sen4,Sen5}. These stable D-branes are basically classified into two classes. One is stable non-BPS D-branes with codimension one \cite{Sen1,Sen3,Sen4,Sen5,BG}. They are stabilized by certain orbifold or orientifold projections, and can be identified with kink solutions. The other is BPS D-branes with codimension two, and they can be interpreted as vortex solutions, or kink solutions on unstable D-branes with codimension one, on the brane -- anti-brane pair \cite{Sen3,Sen5}. Both classes of stable D-branes has been understood in terms of the K-theory in more general way \cite{Witten}, and the author in \cite{Horava} suggested possible applications of this line of arguments to Matrix theory.

In this paper we study the latter mechanism in Matrix theory \cite{BFSS}. We consider a specific example, a system of membrane -- anti-membrane pair,\footnote{Hereafter we will denote membranes and anti-membranes as $M2$ and $\Membar$ respectively. We will also use the abbreviation $Dp$ and $\Dbar{p}$ respectively for D$p$-branes and anti-D$p$-branes.} in which gravitons are supposed to be produced as vortex solutions on the membrane worldvolume via the tachyon condensation:
\begin{equation}
M2 + \Membar \toright{\vev{T}\ne 0} 
                  \mbox{graviton}\,(\mbox{vortex}),
\label{eqn:M2M2bar}
\end{equation}
where $\vev{T}$ denotes the vacuum expectation value of the tachyon fields.

\noindent
This corresponds to the mechanism in type IIA theory \cite{Sen3},
\begin{equation}
D2 + \Dbar{2} \toright{\vev{T}\ne 0} D0\,(\mbox{vortex})\quad
              \mbox{or}\quad\Dbar{0}\,(\mbox{anti-vortex}).
\label{eqn:D2D2bar}
\end{equation}
In Matrix theory we should not expect the appearance of anti-gravitons (corresponding to $\Dbar{0}$ in type IIA theory), for the matrix description of M-theory  is thought of as an infinite momentum limit or as a light-like limit of M-theory. In fact we will see below that we cannot have an anti-vortex solution in the matrix calculation.

\section{The setup}

In string theory the tachyon potential can in principle be computed, but it is hardly possible to carry it out in all order. In Matrix theory, however, we in fact have a simple tachyon potential of quartic order, due to having taken a special limit of M-theory \cite{SS}. Now let us see how tachyon modes appear in the background of the $M2$-$\Membar$ pair in Matrix theory, which was studied in detail in \cite{AB,LM}. The matrix theory action is
\begin{equation}
S = {1 \over R^3M_p^6}\int dt\Tr\left\{
   \half(D_t X^i)^2 +\quarter [X^i,X^j]^2 + \Psi^T D_t\Psi
   +i\Psi^T\gamma^i [X_i,\Psi]\right\},
\label{eqn:Maction}
\end{equation}
where $X^i$ ($i=1,\cdots,9$) and $\Psi$ are $N\times N$ hermitian matrices. $R$ is the radius of a light-like circle and $M_p$ is the eleven dimensional Planck mass. We use a convention $\del_t -i[A_0, \cdot]$ for the covariant derivative $D_t$.

\noindent
We can express the $M2$-$\Membar$ pair spreading over the ($8$,$9$)-directions of interval $b$ in the $7$-direction by the background matrices
\begin{equation}
B^8 = \tbtmtrx{P_1}{0}{0}{P_2},\quad
B^9 = \tbtmtrx{Q_1}{0}{0}{-Q_2},\quad
B^7 = \tbtmtrx{0}{0}{0}{b},
\label{eqn:M2aM2bgd}
\end{equation}
where $P_i$ and $Q_i$ satisfy the commutation relation $[P_i,Q_i]=-ic$ with a constant $c$ which is a charge density of the membrane \cite{BSS}. The upper diagonal block corresponds to $M2$ and the lower to $\Membar$. The background matrices for the other directions, the gauge potential $A_0$ and the fermions are vanishing. For later use we shall represent $P_i$ and $Q_i$ by
\begin{equation}
P_i = c\left(-i\del_{x_i} + A_{x_i}\right), \quad
Q_i = c\left(-i\del_{y_i} + A_{y_i}\right), \\
\label{eqn:CD}
\end{equation}
with the field strength $F_{x_i y_i}=\del_{x_i}A_{y_i}-\del_{y_i}A_{x_i}={1 \over c}$ \cite{KVK}. 
In this representation the trace in the action is replaced with integrals over $x_i$ and $y_i$. For our purpose, we are only interested in the type of a trace
\begin{equation}
\Tr\left[\tbtmtrx{D_{x_1}}{X}{X^{\dagger}}{D_{x_2}}, 
          \tbtmtrx{D_{y_1}}{Y}{Y^{\dagger}}{D_{y_2}}
   \right]^2,
\end{equation}
where $D_{x_i}$ and $D_{y_i}$ denote, in the representation given in eq. (\ref{eqn:CD}), covariant derivatives in which the gauge fields depend only on $x_1$ and $y_1$ for $i=1$, and $x_2$ and $y_2$ for $i=2$. We shall assume that diagonal elements, even the ones obtained as the products of matrices, are local with respect to the coordinates $x_1$ and $y_1$ for the upper block and $x_2$ and $y_2$ for the lower block, although they are in general bi-local. We will factorize a trace as $\Tr = \Tr^{\prime}\otimes\mbox{tr}_{2\times 2}$, and concentrate on a piece $\Tr^{\prime}$. There are three types of traces we have to consider.

\noindent
(i) A trace containing only diagonal elements, i.e., covariant derivatives:
\begin{equation}
\Tr^{\prime}[D_{x_i},D_{y_i}]^2 =\sigma_0
\int dx_i dy_i [D_{x_i},D_{y_i}]^2,
\end{equation}
where $\sigma_0=\frac{1}{2\pi c}$ is a normalization factor and
denotes the density of 
$D0$-branes on the (anti-)$D2$-brane worldvolume in the type IIA
picture \cite{KVK}.

\noindent
(ii) Two diagonal and two off-diagonal elements. An example is
\begin{equation}
\Tr^{\prime}D_{x_1}Y
       D_{y_2}X^{\dagger}
 =\sigma_0^2
  \int dx_1 dy_1 dx_2 dy_2 
     D_{x_1}Y(x_1,y_1;x_2,y_2)
       D_{y_2}X^{\ast}(x_1,y_1;x_2,y_2),
\end{equation}
where $\ast$ denotes the complex conjugation, and the same rule is applied to the other cases of this type.

\noindent
(iii) Four off-diagonal elements. An example is 
\begin{eqnarray}
\Tr^{\prime}XY^{\dagger}XY^{\dagger}
  &=&\sigma_0^3
   \int dx_1 dy_1 dx_2 dy_2 
   dx^{\prime}_2 dy^{\prime}_2
   X(x_1,y_1;x_2,y_2)Y^{\ast}(x_1,y_1;x_2,y_2)
 \nn\\
 && \qquad\qquad\qquad\qquad\qquad\qquad\times
   X(x_1,y_1;x^{\prime}_2,y^{\prime}_2)
   Y^{\ast}(x_1,y_1;x^{\prime}_2,y^{\prime}_2),
\end{eqnarray}
and similar results hold for the other cases.

In order to see the appearance of tachyon modes in this system, it is instructive to compute the potential between $M2$ and $\Membar$ in the one-loop approximation. It can be done most easily in the background gauge with the gauge fixing term
\begin{equation}
S_{GF} = -{1 \over R^3M_p^6}\int dt \half\Tr\left(\del_t A_0 + i[B_i,Y^i]
         \right)^2.
\label{eqn:GF}
\end{equation}
Here the matrices $Y^i$ denotes the fluctuations around the background, that is, $X^i = B^i + Y^i$. 
Since we know that the tachyon modes would come from the open strings stretched between $D2$ and $\Dbar{2}$ in the corresponding type IIA picture, it suffices to consider the off-diagonal elements of the fluctuation matrices (and ghosts  as well):
\begin{equation}
Y^i = \tbtmtrx{0}{\Phi^i}{(\Phi^i)^{\dagger}}{0}, \quad
A_0 = \tbtmtrx{0}{\phi}{(\phi)^{\dagger}}{0}, \quad
\Psi = \tbtmtrx{0}{\psi}{(\psi)^{T}}{0}.
\label{eqn:offdiag}
\end{equation}
Then the mass matrices for the off-diagonal fluctuations are given by
\begin{eqnarray}
&M^2_{\Phi^i} = -M^2_{\phi} = M^2_{\mbox{\scriptsize{ghosts}}} 
  = b^2 + H, \quad (i=1,\cdots,7),& \\
&M^2_{(\Phi^8,\Phi^9)}=\tbtmtrx{b^2+H}{4ic}{-4ic}{b^2+H},& \\
&M_{\psi}=-\gamma^7 b +\gamma^8 (P_1- P_2^{\ast})
           +\gamma^9 (Q_1+ Q_2^{\ast}),&
\label{eqn:massmtrx}
\end{eqnarray}
where $H = (P_1- P_2^{\ast})^2 + (Q_1+ Q_2^{\ast})^2$, and it describes the Hamiltonian of a charged particle moving on a two-dimensional plane in the presence of a uniform magnetic field perpendicular to the plane. Thus the energy level of the Hamiltonian $H$ is identical to the Landau level, $4c(n+\half)$ ($n=0,1,\cdots$).

Now we can find that the tachyon develops when $M2$ and $\Membar$ approach to some critical interval, by diagonalizing the mass matrices for ($8,9$)-directions:
\begin{equation}
U^{\dagger}M^2_{(\Phi^8,\Phi^9)}U = \tbtmtrx{b^2+2c(2n+3)}{0}{0}{b^2+2c(2n-1)},
\qquad U={1 \over \sqrt{2}}\tbtmtrx{i}{-i}{1}{1}.
\label{eqn:diagmass}
\end{equation}
The upper eigenvalue is the mass square of the matrix $\bar{\Phi}\equiv{1 \over \sqrt{2}}(\Phi^9+i\Phi^8)$, and the lower is that of $\Phi\equiv{1 \over \sqrt{2}}(\Phi^9-i\Phi^8)$. The mass square of $\Phi$ becomes negative for the ground state when the interval $b$ is shorter than $\sqrt{2c}$, and thus the ground state of $\Phi$ corresponds to a complex tachyon mode in the $D2$-$\Dbar{2}$ pair for $b < \sqrt{2c}$ \cite{AB,LM}.

We can easily carry out the one-loop calculation, and the long range potential between $M2$ and $\Membar$ are reliably calculated in this approximation. The result is
\begin{equation}
V(b) \sim -N\int_0^{\infty}{ds \over s}s^{-1/2}e^{-b^2s}
        \frac{\sinh^4 cs}{2\sinh 2cs}
  \,\,\toright{b\to\infty}\,\, -\quarter\Gamma(5/2){Nc^3 \over b^5},
\label{eqn:longpot}
\end{equation}
and the agreement with the supergravity result was found at least up to a numerical factor \cite{AB,LM}. Speaking repeatedly, the potential (\ref{eqn:longpot}) diverges for $b < \sqrt{2c}$, and it exhibits the tachyon instability of this system. A lesson from the one-loop computation is that the tachyon mode appears in the fluctuation $\Phi$ when the distance between $M2$ and $\Membar$ becomes shorter than the critical value $\sqrt{2c}$.

\section{The effective theory of the $M2$-$\Membar$ system}

We are now interested in the region $b \le \sqrt{2c}$. At the critical distance $b=\sqrt{2c}$, a massless mode comes out and it might be related to the horizon of a black hole \cite{KL}. We will mention this point later. For the region $b < \sqrt{2c}$, a complex tachyon develops and thus it would annihilate the $M2$-$\Membar$ pair by its condensation. Moreover, as mentioned before, gravitons are supposed to be produced in this process as vortices on the membrane worldvolume. Let us discuss this mechanism in the case of the coincident $M2$-$\Membar$ pair, that is, $b=0$ case. For this purpose, it is enough to consider only the matrices $X^8$ and $X^9$, and to set the others to zero. The action in our concern is 
\begin{equation}
S_{(8,9)}={1 \over R^3M_p^6}\int dt\half\Tr[X^8,X^9]^2.
\label{eqn:action89}
\end{equation}
The matrices $X^8$ and $X^9$ take the form
\begin{equation}
X^8 = \tbtmtrx{P_1 +ca_{x_1}}{\Phi^8}{(\Phi^8)^{\dagger}}{P_2 +ca_{x_2}},
\quad
X^9 = \tbtmtrx{Q_1 +ca_{y_1}}{\Phi^9}{(\Phi^9)^{\dagger}}{-Q_2 -ca_{y_2}},
\label{eqn:X8X9}
\end{equation}
where $a_{x_i}$ and $a_{y_i}$ denote the diagonal fluctuations.
It is straightforward to show that the action (\ref{eqn:action89}) can be written, after setting $\bar{\Phi}=0$ (defined below the eq. (\ref{eqn:diagmass})), as
\begin{eqnarray}
S_{(8,9)}&=&-{1 \over R^3M_p^6}\int dt\Biggl[\sigma_0^2c^2\int
 dx_1dy_1dx_2dy_2
 \left\{\half |D_z\Phi |^2 -F_{xy}|\Phi|^2\right\}\nn\\
 &&\qquad\qquad
+ \sigma_0c^4\int dx_1dy_1 {1 \over 8}(F_{xy}+ \tilde{F}_{xy})^2
   + \sigma_0c^4\int dx_2dy_2 {1 \over 8}(F_{xy}- \tilde{F}_{xy})^2
    \nn\\
&&  + {\sigma_0^3 \over 2}\int dx_1dy_1dx_2dy_2 dx^{\prime}_2 dy^{\prime}_2 
  \Phi(x_1,y_1;x_2,y_2)\Phi^{\ast}(x_1,y_1;x_2,y_2)\nn\\
&&\qquad\qquad\qquad\qquad\qquad\qquad\qquad\quad\times
  \Phi(x_1,y_1;x^{\prime}_2,y^{\prime}_2)
   \Phi^{\ast}(x_1,y_1;x^{\prime}_2,y^{\prime}_2)\nn\\
&&  + {\sigma_0^3 \over 2}\int dx_2dy_2 dx_1dy_1 dx^{\prime}_1 dy^{\prime}_1 
  \Phi^{\ast}(x_1,y_1;x_2,y_2)\Phi(x_1,y_1;x_2,y_2)\nn\\
&&\qquad\qquad\qquad\qquad\qquad\qquad\qquad\quad\times
  \Phi^{\ast}(x^{\prime}_1,y^{\prime}_1;x_2,y_2)
  \Phi(x^{\prime}_1,y^{\prime}_1;x_2,y_2)\Biggr].
\label{eqn:AH}
\end{eqnarray}
The covariant derivative $D_{z}$ is defined by 
\begin{eqnarray}
D_z&=&D_y +iD_x, \\
D_x&=&\del_x + iA_x +ia_x=(\del_{x_1}+\del_{x_2})
+i(A_{x_1}-A_{x_2}) +i(a_{x_1}-a_{x_2}), \\
D_y&=&\del_y +iA_y +ia_y=(\del_{y_1}-\del_{y_2})
+i(A_{y_1}+A_{y_2})+i(a_{y_1}+a_{y_2}),
\end{eqnarray}
and the field strengthes $F_{xy}$ and $\tilde{F}_{xy}$ by
\begin{eqnarray}
F_{xy} &=& -i[D_x,D_y],\quad
\tilde{F}_{xy}= -i[\tilde{D}_x,\tilde{D}_y],\\
\tilde{D}_x&=&\del_x +i\tilde{A}_x + i\tilde{a}_x 
=\del_x +i(A_{x_1}+A_{x_2})+i(a_{x_1}+a_{x_2}),\\
\tilde{D}_y&=&\del_y +i\tilde{A}_y + i\tilde{a}_y 
=\del_y +i(A_{y_1}-A_{y_2})+i(a_{y_1}-a_{y_2}).
\end{eqnarray}
We shall restrict the configuration of a scalar field $\Phi$ into the
form, $\Phi (x_1,y_1;x_2,y_2)=\Phi (x,y)\sqrt{\delta
(\sqrt{\sigma_0}\tilde{x})\delta 
(\sqrt{\sigma_0}\tilde{y})}$,  where $\tilde{x}\equiv \half (x_1 -x_2)$ and
$\tilde{y}\equiv \half(y_1 + y_2)$ are the center of mass coordinates
of the $M2$ and $\Membar$ system. Then the action (\ref{eqn:AH})
reduces to 
\begin{eqnarray}
S_{(8,9)}&=&-{4\sigma_0 \over R^3M_p^6}\int dt\int dx dy
 \left[{c^2 \over 2} |D_z\Phi |^2 -c^2 F_{xy}|\Phi|^2 +|\Phi|^4 
 +{c^4 \over 4} F_{xy}^2 + {c^4 \over 4}
  \tilde{F}_{xy}^2\right]
\label{eqn:u2gauge}\\
&=&-{4\sigma_0 \over R^3M_p^6}\int dt\int dx dy
 \left[{c^2 \over 2} |D_z\Phi |^2
 +\left\{\left(|\Phi|^2-c\right)^2-c^2\right\}
 +c^2\right. \nn\\
&&\left.\qquad\qquad\qquad\qquad\qquad\qquad\qquad\qquad
-c^2 f_{xy}|\Phi|^2 +{c^4 \over 4} f_{xy}^2 +{c^4 \over 4}\tilde{f}_{xy}^2
 \right],
\label{eqn:tacpot}
\end{eqnarray}
where $f_{xy}=\del_x a_y -\del_y a_x(=F_{xy}-{2 \over c})$ and $\tilde{f}_{xy}=\del_x\tilde{a}_y -\del_y\tilde{a}_x(=\tilde{F}_{xy})$, and they are field strengthes for diagonal fluctuations.
When the tachyon field $\Phi$ is at the lowest Landau level and the field strengthes $f_{xy}$ and $\tilde{f}_{xy}$ for diagonal fluctuations are vanishing, the action (\ref{eqn:tacpot}) reduces to the sum of the tachyon potential $\left\{\left(|\Phi|^2-c\right)^2-c^2\right\}$ and the total amount $c^2$ of tensions of $M2$ and $\Membar$. We can further rewrite the action (\ref{eqn:u2gauge}) as
\begin{eqnarray}
S_{(8,9)}=-{4\sigma_0 \over R^3M_p^6}\int dt\int dx dy
 \left[{c^2 \over 2} |D_z \Phi |^2 +
\left( {c^2 \over 2} F_{xy} - |\Phi |^2\right)^2
+{c^4 \over 4} \tilde{F}_{xy}^2\right],
\label{eqn:Bog}
\end{eqnarray}
and this provides us with a Bogomol'nyi-type equation \cite{Bogomol},
\begin{eqnarray}
D_z \Phi = 0, \quad
{c^2 \over 2} F_{xy} - |\Phi |^2 = 0,\quad
\tilde{F}_{xy} = 0.
\label{eqn:bnd}
\end{eqnarray}
The effective theory (\ref{eqn:Bog}) of the $M2$-$\Membar$ system is a $U(2)$ gauge theory with the background of a constant magnetic field, discussed by \cite{GNS,Sen2}.

At this point we would like to note that our membrane has a topology
of two-torus \cite{BFSS}. The periodicity of the (anti-)membrane is
given by $0\le P_i <\sqrt{\pi cN}$ and $0\le Q_i <\sqrt{\pi cN}$
($i=1,2$) \cite{AB}. In terms of our specific representation
(\ref{eqn:CD}) of $P_i$ and $Q_i$, the periodicity can be read off
from the formulae $\Tr_{N/2}I_{N/2}=\sigma_0\int dx_1 dy_1I_{N/2}$ and
$\Tr_{N/2}I_{N/2}=\sigma_0\int dx_2 dy_2I_{N/2}$, and the
area $A_{T^2}$ of torus, 
spanned by these coordinates, is equal to $\pi cN$, i.e., the same as
that of ($P_i,Q_i$)-planes. Let the periodicities of $x$ and $y$ be
$0\le x<R_x$ and $0\le y<R_y$ respectively. Then $R_x$ and $R_y$
satisfy $R_xR_y=A_{T^2}=\pi cN$. We would like to emphasize that the
effective theory obtained in (\ref{eqn:Bog}) is not the Matrix theory
compactified on $T^2$. Our interpretation of the effective theory is
similar to the one discussed in \cite{KVK}, in which it is justified
that the worldvolume theory obtained via the representation
(\ref{eqn:CD}) can be thought of as that on the membrane itself, not
on the T-dualized torus.

We would also like to remark that there is obviously a decoupled $U(1)$ gauge field in the effective theory. As pointed out in \cite{Sred,Witten}, even the decoupled $U(1)$ here should be eliminated in the $M2$-$\Membar$ annihilation. The author of \cite{Yi} proposed a resolution of this puzzle. It is not the same mechanism, but similarly we would like to point out that the decoupled $U(1)$ might be confined, that is, there is a mass gap  and thus no massless vector in this $U(1)$ theory, if we can justify that this $U(1)$ is compact and the Polyakov's argument \cite{Polyakov} goes through as well in the case of a two-torus.

The background magnetic field gives a 't Hooft's twisted boundary condition \cite{GNS,Sen2}. We shall take the following form of the background fields:
\begin{eqnarray}
{\cal A}_x=\tbtmtrx{A_x}{0}{0}{-A_x}=-{1 \over c}y\sigma_3,
\quad
{\cal A}_y=\tbtmtrx{A_y}{0}{0}{-A_y}=0.
\label{eqn:bcgrnd}
\end{eqnarray}
Then the boundary conditions for these fields are given by
\begin{eqnarray}
{\cal A}_{\mu}(R_x,y)&=&{1 \over i}\Omega_x\del_{\mu}
\Omega_x^{-1}+\Omega_x{\cal A}_{\mu}(0,y)\Omega_x^{-1},\nn\\
{\cal A}_{\mu}(x,R_y)&=&{1 \over i}\Omega_y\del_{\mu}
\Omega_y^{-1}+\Omega_y{\cal A}_{\mu}(x,0)\Omega_y^{-1},
\label{eqn:twstbc}
\end{eqnarray}
where transition functions $\Omega_x$ and $\Omega_y$ are 
\begin{equation}
\Omega_x(x,y)=I_2,\quad
\Omega_y(x,y)=e^{iR_y x\sigma_3/c},
\label{eqn:omega}
\end{equation}
and they satisfy a consistency condition given by 't Hooft \cite{tHooft}
\begin{equation}
\Omega_x(y=0)\Omega_y(x=R_x)
=\Omega_y(x=0)\Omega_x(y=R_y)(-1)^N.
\label{eqn:consistency}
\end{equation}
Corresponding to these boundary conditions, we must have the boundary conditions for the other fields via
\begin{eqnarray}
X^{i=8,9}(R_x,y)=\Omega_xX^{i=8,9}(0,y)\Omega_x^{-1},
\quad
X^{i=8,9}(x,R_y)=\Omega_yX^{i=8,9}(x,0)\Omega_y^{-1},
\label{eqn:bndoffdiag}
\end{eqnarray}
and we can find that these conditions amount to
\begin{equation}
\left\{
 \begin{array}{c}
 \Phi(R_x,y)=\Phi(0,y)\\
 \bar{\Phi}(R_x,y)=\bar{\Phi}(0,y)
 \end{array},
\right.
\qquad
\left\{
 \begin{array}{c}
 \Phi(x,R_y)=e^{2iR_yx/c}\Phi(x,0)\\
 \bar{\Phi}(x,R_y)=e^{2iR_yx/c}\bar{\Phi}(x,0)
 \end{array}.
\right.
\label{eqn:fluxbc}
\end{equation}

\section{Tachyon condensation and graviton production}

Now we shall show that there exist vortex-like solutions of the
Bogomol'nyi-type equation (\ref{eqn:bnd}). Let us express the `Higgs'
field $\Phi$ by two real valued functions $u$ and $\Theta$ as $\Phi =
\exp[\half(u+i\Theta)]$. From the first equation in (\ref{eqn:bnd})
the $U(1)$ gauge field $A_z +a_z=\half\left[(A_x +a_x)-i(A_y
+a_y)\right]$ is described by $A_z +a_z = {i \over
2}\del_z(u+i\Theta)$, where $\del_z =\half(\del_x -i\del_y)$. Set the
angular part $\Theta$ of the `Higgs' field as $\Theta =
i\sum_{k=1}^n\left[\log\vartheta_1({z-z_k \over
2\pi}|\tau)-\log\overline{\vartheta_1({z-z_k \over
2\pi}|\tau)}\right]$, where the overline denotes the complex
conjugation and $\Theta$ is a multivalued function.\footnote{On ${\bf
R}^2$, $\Theta$ may be chosen as $-2\sum_{k=1}^n \mbox{arg}(z-z_k)$.}
Then the second equation in (\ref{eqn:bnd}) reduces to 
\begin{equation}
c^2\del_z\del_{\bar{z}}u = -e^u +2\pi c^2\sum_{k=1}^{n}\delta^2(z-z_k).
\label{eqn:NLPDEQ}
\end{equation}
This is a Liouville-type equation discussed by Jackiw and Pi \cite{JP}, and general multi-vortex solutions are constructed at least on ${\bf R}^2$ \cite{JP} and an explicit one-vortex solution on $T^2$ is given in \cite{Olesen}. We, however, do not follow their constructions and adopt another strategy in order to handle the boundary conditions of vortex solutions manifestly.

Now we divide the radial part $u$ of the `Higgs' field $\Phi$ into a singular part $u_0$ and an analytic part $v$. The singular function $u_0$ takes the form
\begin{equation}
u_0 =\sum_{k=1}^n\left\{\log\left|\vartheta_1\left(\left.{z-z_k \over 2\pi}
      \right|\tau
      \right)\right|^2
     -\frac{1}{2\pi\tau_2}\left[\mbox{Im}(z-z_k)\right]^2\right\},
\label{eqn:sing}
\end{equation}
where $\tau_2=\mbox{Im}\tau$ and this is exactly doubly periodic
\cite{Polchinski}. The analytic function $v$ obeys the equation
\begin{equation}
c^2 \del_z\del_{\bar{z}}v =c^2 \frac{n}{4\pi\tau_2}
 -\prod_{k=1}^n\left|\vartheta_1\left(\left.\frac{z-z_k}{2\pi}
  \right|\tau
   \right)\right|^2
 e^{-{1 \over 2\pi\tau_2}\left[{\scriptsize \mbox{Im}}(z-z_k)
    \right]^2}e^v.
\label{eqn:analytic}
\end{equation}
By the theorem 7.2 of Kazdan and Warner \cite{KW}, one can show the
existence of a solution for this equation, at least if
$\frac{n}{4\pi\tau_2}$ is sufficiently small.\footnote{Their
theorem is the following. Consider a differential equation
$\del_z\del_{\bar{z}}v = C - He^v$, on Riemann surfaces, where $C$ is
a positive constant and $H$ is an analytic function. Then there exists
an analytic solution $v$, if and only if $H$ is positive somewhere and
$C<C_{+}(H)$, an upper bound depending on $H$. But in our case we can
expect that $n/4\pi\tau_2$ does not need to be small, since generic solutions of the Liouville equation (\ref{eqn:NLPDEQ}) is known and it ensures very likely the existence of a solution of eq. (\ref{eqn:analytic}).} 

Let us consider the boundary conditions for `Higgs' field $\Phi$. They can be easily read off from the angular part $\Theta$ of the `Higgs' field:
\begin{eqnarray}
\Phi\left(z +2\pi\tau\right)&=&e^{-i\pi n
  +i\pi n{\scriptsize \mbox{Re}}\tau +i\sum_{k=1}^n
  {\scriptsize \mbox{Re}}(z-z_k)}\Phi\left(z\right),\\
\Phi\left(z+2\pi\right)&=&e^{-i\pi n}\Phi
\left(z\right).
\label{eqn:phibc}
\end{eqnarray}
Comparing these with (\ref{eqn:fluxbc}) and noting that $R_x=2\pi$ and
$R_y=2\pi\tau_2=cN/2$ in this convention and that we can obviously
replace $x=\mbox{Re}z$ in (\ref{eqn:omega}) and (\ref{eqn:fluxbc})
with ${1 \over N}\sum_{k=1}^{N}(x-x_k)$, one can find that those
conditions coincide exactly by setting $n=N(=\mbox{even})$.

At this point one may wonder if a smilar analysis goes through even
for the case that $\bar{\Phi}\ne 0$ and $\Phi=0$, that is, regardless
of the existence of the tachyon.\footnote{Throughout this paper we are
assuming $c>0$.} It is almost true. The Bogomol'nyi-type equation
(\ref{eqn:bnd}) becomes $D_{\bar{z}}\bar{\Phi}=0$ where
$D_{\bar{z}}=-iD_x+D_y$, and ${c^2 \over 2}F_{xy}+|\bar{\Phi}|^2=0$,
and thus we can solve it similarly.
There is, however, a significant difference. In this case the angular part
$\Theta$ flips its sign and we obtain the boundary conditions for
$\bar{\Phi}$ with wrong sign. Thus we cannot construct vortex
solutions for $\bar{\Phi}$ with the correct boundary conditions. 

Now let us discuss the tachyon condensation in this system. The effective theory is a pure $U(2)$ gauge theory and thus it seems that the tachyon cannot condense. We can, however, show that it eventually does. Integrating (\ref{eqn:analytic}) over $T^2$, we can rewrite it as
\begin{equation}
c^2\frac{N}{4\pi\tau_2}A_{T^2}=\int_{T^2}dzd\bar{z}|\Phi|^2.
\label{eqn:vev}
\end{equation}
From the above argument, we have $N/4\pi\tau_2= 1/c$. Therefore we can conclude that the average value of $|\Phi|$ is equal to $\sqrt{c}$. It is anticipated by the form of the tachyon potential $\{(|\Phi|^2-c)^2-c^2\}$ which we observed before, and this shows the tachyon condensation in the $M2$-$\Membar$ system.

Now the magnetic field $F_{xy}$ on the worldvolume of the coincident $M2$-$\Membar$ pair is given by $F_{xy}=\frac{N}{2\pi\tau_2} -2\del_z\del_{\bar{z}}v$, and thus the magnetic flux is 
\begin{equation}
\int_{T^2}dzd\bar{z} F_{xy} = 2\pi N\,(=2\pi\cdot 2\sigma_0\cdot A_{T^2}).
\label{eqn:firstChern}
\end{equation}
This corresponds to the $N$-vortex solution on $T^2$, in which each
vortex is sitting at one of the zeros $z_k$ of the `Higgs' field. As
is well-known, a magnetic flux on the membrane worldvolume couples to
a RR one-form via a Chern-Simons coupling and thus gives a $D0$-brane
charge in the corresponding type IIA picture. The number of
$D0$-branes corresponding to the above flux is equal to $N$, since the
value $2\pi N$ of the flux is just twice the one ($\int dx_1dy_1{1 \over
c}=\int dx_2dy_2{1 \over c}=\pi N$) on each (anti-)membrane which is
constituted from $N/2$ $D0$-branes.
Therefore the process we obtained will be 
\begin{equation}
M2\,\, (+\,\, N/2\,\, \mbox{M-mom}) +
\Membar\,\, (+\,\, N/2\,\, \mbox{M-mom})
\,\,\toright{\vev{\Phi}\ne 0}\,\,
\mbox{gravitons}\, (+\,\, N\,\, \mbox{M-mom}),
\label{eqn:label}
\end{equation} 
where ``M-mom" indicates the M-momentum, and this process in fact
conserves the M-momentum. We will justify below that vortices can be
interpreted as gravitons, as denoted in the r.h.s. of the above
equation. 
 
Now let us show that vortices can be thought of as gravitons in Matrix theory.
The bound for the energy is exactly zero \cite{Amb}. This indicates that vortices in this system are massless particle, and thus they can be regarded as gravitons in M-theory. It, however, seems somewhat strange that a massless particle does not carry any energy. But remember that the matrix theory Hamiltonian is of the form $H =({\bf P}_{\perp}^2 +M^2)/2P_-$. Hence in matrix theory the Hamiltonian of a massless particle without transverse momentum is exactly zero, which accords with the above result.  We would like to note that anti-gravitons cannot appear in this system, since as we have shown the magnetic flux is positive. Again it is in accordance with the interpretation of Matrix theory \cite{BFSS}. A graviton of this type carries the M-momentum in the form of a magnetic flux $\int dxdy F_{xy}$, as mentioned above, which may be reminiscent of the membrane scattering with M-momentum transfer \cite{PP}\cite{HKS}. A vortex solution is, however, a static solution, as is different from an instanton solution which can provide scattering processes with M-momentum transfer. Hence a vortex solution cannot describe a transfer of M-momentum.

We would also like to remark that, at the minimum of the tachyon potential, the sum of the tension of $M2$ and $\Membar$ is exactly cancelled by the energy density of the tachyon field, as discussed in \cite{Sen2,Sen3}. We can observe it from the action (\ref{eqn:tacpot}): (here in unit $\sigma_0/R^3M_p^6=1$)
\begin{eqnarray}
&&E_{M2} +  E_{\Membar}=-\half\Tr[B^8,B^9]^2=
\left(\int dx_1dy_1 + \int dx_2 dy_2\right)\half c^2
 =4\int dx dy\, c^2,
 \\
&&E_{\mbox{\scriptsize{tachyon}}}=4\int dx dy
\left\{ (|\Phi |^2-c)^2 -c^2\right\}\toright{|\Phi|=\sqrt{c}}
-4\int dx dy\, c^2.
\end{eqnarray}
This indicates the restoration of supersymmetry \cite{Sen2}.
Now let us discuss the supersymmetry of this system. The initial state, the $M2$-$\Membar$ pair, breaks all of the supersymmetry. As we showed above, it is annihilated via the tachyon condensation and gravitons are produced as vortices on the membrane worldvolume. The final state, gravitons, is a massless state, and thus it must preserve a half of the supersymmetry. Let us look at the fermion transformation law under the supersymmetry transformations:
\begin{equation}
\delta\Psi =\left(D_t X^i \gamma_i -{i \over 2}
   [X^i, X^j]\gamma_{ij}\right)\epsilon +\epsilon^{\prime}.
\label{eqn:SUSY}
\end{equation}
For a vortex solution we can find that 
\begin{eqnarray}
\delta\Psi &=&-{i \over 2}
\tbtmtrx{-i\left[{c^2 \over 2}(F_{xy}+\tilde{F}_{xy}) -|\Phi|^2\right]}
{-{c \over \sqrt{2}}D_z \Phi}
{{c \over \sqrt{2}}(D_z \Phi)^{\ast}}
{i\left[{c^2 \over 2}(F_{xy}-\tilde{F}_{xy}) -|\Phi|^2\right]}\gamma_{89}
\epsilon + \epsilon^{\prime}
\nn\\
&=& \epsilon^{\prime}.
\label{eqn:voSUSY}
\end{eqnarray}
This shows that a vortex solution indeed preserves a half of the supersymmetry, as we expected. Note also that gravitons in Matrix theory are described by a background of commuting matrices. The above result (\ref{eqn:voSUSY}) is nothing but the statement that $X^8$ and $X^9$ are commuting for a vortex solution. Again this agrees with the interpretation of a vortex as a graviton in Matrix theory.

\section{Discussions}

Now we would like to discuss future directions of this work. The
authors in \cite{KL} pointed out the similarity of the tachyon
instability to the absorption of matters by a non-extremal black
hole. They argued that this similarity leads to the identification of
the horizon of a non-extremal black hole with a critical point
(corresponding to a point $b=\sqrt{2c}$ in the $M2$-$\Membar$ system)
where a massless mode comes out. As the horizon has a light-cone
structure, the appearance of a massless mode matches this
feature. Along the line of their arguments, it would be interesting to
pursue the following possibility in the $M2$-$\Membar$ system. We will
indicate the corresponding picture in terms of a black hole in
parenthesis:  

Set $M2$ and $\Membar$ infinitely separated, that is,
$b(t=-\infty)=\infty$. The attractive force between $M2$ and $\Membar$
makes them approach up to a critical point (the horizon),
$b(t)=\sqrt{2c}$, where a massless mode appears, and subsequently they
across this point (the horizon) to exhibit the tachyon
instability. Then there appear vortices (gravitons radiated by a black
hole) which are time-dependent through $b(t)$ and thus carry transverse
momenta in the region, $0 < b(t) < \sqrt{2c}$. We
expect that $M2$ and $\Membar$ gradually collapses to a point,
$b(t=\infty)=0$, and there remain only vortices (gravitons) as discussed
above. In this senario the initial energy of $M2$ 
and $\Membar$ seems to disappear, for a vortex at the final state does
not carry any light-cone energy as mentioned before. But Matrix theory
is unitary and thus the total energy of $M2$ and $\Membar$ must be
radiated (the Hawking radiation) by a certain mechanism in which
vortices (gravitons), in the region $0<b(t)<\sqrt{2c}$, carry away the
initial energy. 

Another direction is to study stable non-BPS states
\cite{Sen1,Sen3,Sen4,Sen5,BG} in Matrix theory. To do this, we need
certain orbifold or orientifold projections to make a complex tachyon
real so that there can exist a kink solution on the brane --
anti-brane pair. As an example it is straightforward to apply our
analysis to a Matrix theory on an orientifold background \cite{OM}. We
will report some progress in this direction elsewhere \cite{AHH}. It
also seems possible to discuss stable non-BPS states in type I theory
\cite{Sen3,Sen4} by making use of the $USp(2k)$ matrix model
\cite{IT}. 

Finally we would like to mention that our analysis is applicable to
more general systems, such as general brane -- anti-brane pairs in the
IIB matrix model \cite{IKKT} as well as those in Matrix theory. 

\section*{Acknowledgments}

We are grateful to M.~Fukuma, Y.~Imamura, H.~Kunitomo, Y.~Matsuo and K.~Murakami for useful discussions. H.~A. would like to thank H.~Itoyama and A.~Tsuchiya for interesting conversations. S.~H. would like to thank A.~Kato for helpful discussions, and Y.~Kazama and T.~Kuroki for useful conversations. The work of S.~H. was supported in part by the Japan Society for the Promotion of Science.
%%%%%%%%%%%%%%%%%%%%%%%%%%%%%%%%%%%%%%%%%%%%%%%%%%%%%%%%%%%%%%%%%%%%%%%%%%%%%%

%References

%%%%%%%%%%%%%%%%%%%%%%%%%%%%%%%%%%%%%%%%%%%%%%%%%%%%%%%%%%%%%%%%%%%%%%%%%%%%%%

\end{document}